\documentclass{iaus}
\usepackage{graphicx}

\title[Weak lensing and gravity theories]{Weak lensing and gravity theories}

\author[V. Acquaviva, C. Baccigalupi and F. Perrotta]{Viviana Acquaviva, Carlo Baccigalupi, Francesca Perrotta}

\affiliation{SISSA/ISAS, Via Beirut 4, 34014 Trieste, Italy, and \\
INFN, Sezione di Trieste, Via Valerio 2, 34127 Trieste, Italy\\
 email:acqua@sissa.it; bacci@sissa.it; perrotta@sissa.it}

\pubyear{2004}
\volume{225}
\pagerange{1--8}
\date{?? and in revised form ??}
\jname{Impact of Gravitational Lensing on Cosmology}
\editors{Mellier, Y. \& Meylan,G. eds.}
\begin{document}

\maketitle

\begin{abstract}
We present the theory of weak gravitational lensing in cosmologies with generalized gravity, described in the Lagrangian by a generic 
function depending on the Ricci scalar and a non-minimally coupled scalar 
field. \\
We work out the generalized 
Poisson equations relating the dynamics of the fluctuating components 
to the two gauge invariant scalar gravitational potentials, fixing the new 
contributions from the modified background expansion and fluctuations. \\
We show how the lensing observables are affected by the cosmic expansion as 
well as by the presence of the anisotropic 
stress, which is non-null at the linear level both in scalar-tensor gravity 
and in theories where the gravitational Lagrangian term features a 
non-minimal dependence on the Ricci scalar. 
We derive the generalized expressions for the convergence power spectrum,
and illustrate phenomenologically the new effects in Extended Quintessence scenarios, 
where the scalar field coupled to gravity plays the role of the dark energy.
\end{abstract}

\firstsection
\section{Introduction}

In the recent years several independent datasets, namely the
distant type Ia supernovae (\cite{riess_etal,perlmutter_etal}), 
the Cosmic Microwave Background (CMB) anisotropies (\cite{bennett_etal} and
references therein), the Large Scale Structure 
(LSS, \cite{percival_etal,dodelson_etal}) 
and the Hubble Space Telescope (HST, \cite{freedman_etal}) have revealed that our Universe is currently undergoing a phase of cosmic acceleration.
The search for an explanation to this unexpected phenomenon has been one of the most interesting research topics of the last years, and the picture is still far from being satisfactory.\\
The simplest description of the vacuum 
energy responsible for cosmic acceleration is a purely geometric term 
in the Einstein equations, the Cosmological Constant. This explanation, though appealing in its simplicity, raises obvious fine-tuning issues, part of which can be solved if the concept of Cosmological Constant is extended to a dynamical vacuum component, commonly referred to as the dark energy 
(see \cite{sahni_starobinski,peebles_ratra,padmanabhan} and references therein). \\
The most straightforward generalization, already introduced well before the
evidence for cosmic acceleration, is a scalar 
field, dynamical and fluctuating, with a background evolution slow enough to
mimic a constant vacuum energy given by its potential, providing 
cosmic acceleration.
In particular, it was demonstrated how the dynamics of this component, 
under suitable potential shapes inspired by super-symmetry and 
super-gravity theories, can possess attractors in the
trajectory space, named tracking solutions, capable to reach the
present dark energy density starting from a wide set of initial 
conditions in the very early universe, thus alleviating, at least
classically, the problem of fine-tuning. Finally, we regard the possibility of cosmic acceleration arising from the gravitational sector of the theory only, implemented as a modification of the action of General Relativity, admitting a 
general dependence on the Ricci and Brans-Dicke scalar fields. \\
We regard as an investigation tool the weak lensing shear, which was 
detected recently by independent groups with 
astonishing agreement (\cite{bacon_etal},\cite{wilson_etal},\cite{wittmann_etal},
\cite{maoli_etal},\cite{vanwaerbecke_etal}). Although the precision of such 
measurements does not allow to constrain different cosmological 
models, the planned observations will become certainly 
a crucial tool to investigate the behavior of dark matter and energy. 
The reason of this interest is the timing: the structure formation, and 
the weak lensing carrying its physical information, occurs at an epoch 
which overlaps with the onset of cosmic acceleration; by virtue of 
this fact, it is reasonable to expect a good sensitivity of the weak 
lensing effect to the main dark energy properties such as the equation 
of state and its redshift behavior. \\
We follow the harmonic approach to weak lensing (\cite{hu}) and the 
treatment for generalized cosmological scenarios including 
cosmological perturbations (\cite{hwang}), already exploited for 
investigating the effects of the explicit coupling between dark 
energy and gravity (\cite{matarrese_baccigalupi_etal} and 
references therein). \\
This work is organized as follows. We take advantage of the previously developed (\cite{acqua_etal}) formalism for the treatment of cosmological perturbations in generalized scenarios, thus including all the above mentioned explanations for the cosmic acceleration, be they a cosmological constant, a dynamical scalar field or rather a modification of the Einstein formulation of general relativity. The physical features imprinted by the weak lensing on the structure on the CMB multipoles, which are the ultimate object of our analysis, are briefly reviewed in the next section; we will then move on to the description of the major new effects arising in these scenarios with respect to the case of ordinary cosmology, and we provide a numerical flavor of the amplitude and the size of the modifications in the specific model of Extended Quintessence (\cite{perrotta_baccigalupi_etal}). Finally, we give an insight of the next steps towards the complete and fully accurate numerical understanding of these features.

\section{Lensing on the CMB signal}

The effect of weak lensing on the CMB multipoles in total intensity in ordinary cosmology has been widely studied and is now fully understood at all scales 
(see e.g. \cite{hu} and references therein).
Very briefly, the way it affects the temperature anisotropies can be summarized as follows:
\begin{itemize} 
\item On the very large scales, the lensing effect is negligible. This is easily understood considering that on scales much larger than the typical size of the lenses, of order of tens of Mpc, the distortions induced by the lensing signal are naturally randomized, and thus the lensing effect is not coherent and the power in the low multipoles is unaffected. 
\item On the intermediate scales, there is a typical length scale, which is quite model-independently set at $\theta \simeq 20'$, corresponding to multipoles of several hundreds, over which the lensing signal is not any more randomized and the net effect of lensing is clearly recognizable as a smoothing of the peaks and troughs structure.    
\item Finally, on very small scales ($l > 2000$), it has been shown (\cite{metcalf_silk}) that the lensed signal sistematically falls 
above the unlensed one, because of the lensing capability to transfer power from smaller multipoles;
thus, an enhancement of power in the featureless damping tail can be safely be regarded as a lensing-induced effect.
\end{itemize} 
Even more interesting, at least as far as our analysis is concerned, is the effect of lensing on the polarization modes of the CMB. 
This is because of a typical lensing property, which is the mixing of electric (E) and magnetic (B) modes 
of polarization (\cite{hu}). Therefore, even if one starts with an absent or negligible contribution from primordial tensor modes, 
some of the power stored in the E 
ones is transferred by lensing, and one ends up with a significant fraction of power converted in B
modes polarization. Furthermore, the signal in the B modes due to lensing simply traces the E power, and 
turns out to be peaked at multipoles $l \simeq 1000$, resulting in its being easily distinguishable from the contribution due to gravitational waves, peaking at much 
larger scales. As a result, the signal encoded in the magnetic modes of polarization can be used as a tracer of the lensing effect, and it effectively is what we regard as the most promising tool for discriminating among different 
cosmological models, especially for what concerns their capability of imprinting cosmic acceleration. 

\section{From lensing to cosmology observables}

Since our aim is to use the lensing observables in order to probe the underlying cosmological model, we need first to identify the path linking the informations encoded in the lensing signal, namely in the distortion tensor $\psi_{ij}$, to the three-dimensional power spectrum of matter.\\
In particular, we know that in the weak lensing hypothesis all the relevant physical information is carried by two observables only, the \textit{convergence} $\kappa$, which describes the isotropic magnification of an image, and the \textit{shear} $\gamma$, which correspondes to its anisotropic distortion. However, since in the weak lensing regime these two quantities can be shown to have identical statistical properties, we will use the power spectrum of the convergence as representative of the lensing features. \\
The relation between this power spectrum and that of the gravitational potential will be given by the equation of motion for photons, namely, the null geodesic equation in the perturbed Friedmann-Robertson-Walker universe:
   
\begin{equation} 
\label{ge}
\frac{d^2 r^\alpha}{d\lambda^2} = -
g^{\alpha\beta}\left(g_{\beta\nu,\mu} -
\frac{1}{2}g_{\mu\nu,\beta}\right) \frac{dr^\mu}{d
\lambda}\frac{dr^\nu}{d\lambda}\,, 
\end{equation}

Once this equation has been solved, we can use the Poisson equation in order to establish the relation between the power spectrum of the gravitational potential $\Phi$, which is the only relevant physical degree of freedom in ordinary cosmologies, and the one of the matter distribution; for a flat $\Lambda$CDM model this is given in the Fourier space by

\begin{equation}
\Delta_{\Phi}^2 = \frac{9}{4}\left(\frac{H_0}{k}\right)^4 \left(1 + 3\frac{H_0^2}{k^2}\Omega_{K}\right)^{-2}\Omega^2_m(1+z)^2\Delta^2_{\delta}\,.
\end{equation}

We will discuss how these fundamental treatment needs to be generalized in order to achieve the same analytical expressions in the modified theories of gravity of interest.

\section{From ordinary to generalized cosmologies}

\subsection{Cosmological setting}

We consider a class of theories of gravity whose action is 
written in natural units as
\begin{equation}
\label{action}
S = \int d^4x \sqrt{- g}\left[ \frac{1}{2\kappa}
f(\phi,R) - \frac{1}{2} \omega(\phi)\phi^{;\mu}\phi_{;\mu} -
V(\phi) + {\cal{L}}_{\rm{fluid}}\right],
\end{equation}
where $g$ is the determinant of the background metric,
$R$ is the Ricci scalar, $\omega$ generalizes the kinetic term,
and ${\cal{L}}_{\rm{fluid}}$ includes contributions from the matter and
radiation cosmological components; $\kappa=8\pi G_{*}$ plays the role
of the ``bare'' gravitational constant, and 
the usual gravity term $R/16\pi G$ has been replaced by the generic 
function $f/2\kappa$ (\cite{hwang,perrotta_baccigalupi_etal}).
Here as throughout the paper Greek indices run from 0 to 3, Latin
indices from 1 to 3.\\
We assume the validity of the linear cosmological perturbations treatment, 
and for the sake of simplicity we consider a flat background geometry. Moreover, we have decided
to deal with scalar-type kind of perturbations only, since the widely most important effect of lensing on the B-modes is given by the transfer of power from primordial electric (ie scalar-type) modes. \\
Therefore, our spacetime line element will be given by

\begin{equation}
\label{lineelement}
ds^2 = a^2\left(-(1 + 2\,\Psi)d\tau^2 + (1 + 2\,\Phi)dl^2\right)\,,
\end{equation}

where $\tau$ is the conformal time and $dl^2 = dr^2 + r^2 d\Omega^2$. \\

\subsection{New features in generalized cosmologies}

The effects arising on the lensing signal due to the dark energy or modified gravity dynamics are present both at the background and at 
the perturbations level. The main effect on the background dynamics is the modification of the measures of distances (propagating 
through all the background variables, such as the scale factor or the Hubble parameter), due to the fact that we allow the strength 
of the gravitational field to be time-varying. For the perturbed quantities there are two main different phenomena: the first is that the gravitational potential $\Phi$ gets modifications from its interactions with the fluctuations in the scalar field $\delta\phi$, thus its numerical values will change with respect to the ordinary case; the second is that the moduli of the two gravitational potentials appearing in the line element 
(\ref{lineelement}) 
are no longer equal, signaling the presence of an extra degree of freedom, usually referred to as anisotropic stress. In particular, if we define the function
\begin{equation}
F = \frac{1}{\kappa}\frac{\partial f}{\partial R}\,,
\end{equation}
we can recast the relation between $\Phi$ and $\Psi$ in the simple form
\begin{equation}
\Phi + \Psi = - \frac{\delta F}{F}\,.
\end{equation}
 
\subsection{Calculation of the convergence power spectrum}

From the general solution of eq. (\ref{ge})
we can compute the expression for the convergence:

\begin{equation}
\kappa = \frac{1}{2}(\psi_{11} + \psi_{22}) = \frac{1}{2}\int^{\chi_{\infty}}_0 d\chi \,f(\chi)\partial^i\partial_i [\Psi(\hat{n},\chi) - \Phi(\hat{n},\chi)]\,,
\end{equation}

where the function $f(\chi) = \chi \,\int d\chi^{\prime}(\chi - \chi^{\prime})/{\chi^{\prime}}\,g(\chi^{\prime})$ describes the background geometry and $g(\chi)$ is the source distribution, which in the case of lensing on the CMB is simply a delta function at the last scattering surface. \\
Taking advantage of the hypothesis of linearity, stating that different modes in the Fourier space are dynamically independent, we get the general expression for the convergence power spectrum as

\begin{eqnarray}
P_{\kappa}(l) & = &\frac{8}{\pi}\int_{0}^{\chi_\infty} d\chi\, f(\chi)
\int_{0}^{\chi_\infty} d\chi^{\prime} f(\chi^{\prime})  
\int \,dk\,k^6\, j_l(k\,\chi) j_l (k \,\chi^{\prime}) \cdot\nonumber\\
&\cdot&\left[\frac{1}{4}\langle {\Psi}(k,\chi ){\Psi}(k,\chi' ) \rangle +
\frac{1}{4}\langle {\Phi}(k,\chi ){\Phi}(k,\chi' )\rangle
-\frac{1}{2}\langle {\Psi}(k,\chi){\Phi}(k,\chi')\rangle \right]\,.
\label{pconvergence}
\end{eqnarray}

The second step is given by the generalization of the Poisson equation; we can compute the correction to the Hubble parameter due to generalized cosmologies, labelled as $H_{gc}$, to be 
\begin{equation}
H^2 = H^2_{\rm{fluid}}+ H^2_{\rm{gc}}; \quad H^2_{\rm{gc}} =\frac{1}{F}\left[ \frac{\omega}{2\,a^2}{\phi^{\prime}}^2 + \frac{RF - f/\kappa}{2} + V - \frac{3{\cal H}F^{\prime}}{a^2}\right]
\end{equation}  
and we finally get 

\begin{eqnarray}
\label{gpe} && P_{{\Phi}} =
\frac{9}{4}\left(\frac{H_0}{k}\right)^4 \left[
\frac{F_{0}}{F}\Omega_{0m} (1 + z) + 
\frac{F_{0}}{F}\Omega_{0r} (1 + z)^{2} +
\frac{1}{(1+z)^{2}}\Omega_{\rm gc}\right]^2
P_\Delta\,.
\end{eqnarray} 

where $\Omega_{\rm gc} = H^2_{\rm{gc}}/H^2_0 $ and $\Delta$ is the gauge-invariant density fluctuation (\cite{kodama_sasaki}).

\section{A numerical example: a Non-Minimally-Coupled model}

In order to get an insight to the corrections in the CMB spectra that we can expect
in this kind of models, we will now work out semi-analytical expressions for the convergence power spectrum 
in a specific model. We define as Non-Minimally-Coupled models those where $f(\phi,R)$ is a linear
function of the Ricci scalar:
\begin{equation}
f/\kappa = F\,R
\end{equation}
and we will assume that the function $F(\phi)$ can be written as
\begin{equation}
\label{mg}
F(\phi) = \frac{1}{8\,\pi\,G} + \xi\,(\phi^2 - \phi^2_0)\,, 
\end{equation}
where $\phi_0$ is the present value of the quintessential scalar field and $\xi$ is a coupling parameter. Notice 
that in this scenario the popular Jordan-Brans-Dicke parameter $\omega_{JBD}$ is $1/32\pi G\phi_{0}^{2}\xi^2 $. \\
In these models the effects of anisotropic stress, proportional to the fluctuations $\delta\phi$, can be shown to
be negligible with respect to the background dynamics modifications, due to the time-varying gravitational constant; 
as an example we give the modified radial coordinate
\begin{equation}
\delta r = 4\pi G\xi\phi^2_0\int^z_0\frac{dz}{H(z)}\left(\frac{\phi^2}{\phi^2_0} - 1\right)
\end{equation}
and we notice that, since in these models the trajectory of the field is always a monotonic and increasing function of the redshift, the effective gravitational constant is increasing with time, as seen from eq. 
(\ref{mg}), 
and the universe is shrinking in response, as one would expect.\\
 The resulting correction to the convergence power spectrum is:

\begin{eqnarray}
\delta P _{{\kappa}}(l)& =& -128G\xi\phi_{0}^{2}
\int_{0}^{\chi_\infty} d\chi\, f(\chi)
\int_{0}^{\chi_\infty} d\chi^{\prime} f(\chi^{\prime})
\left(\frac{\phi^{2}}{\phi_{0}^{2}}-1\right) \cdot \\ 
\label{EQ:Pk}
& & \int \,dk\,k^6\, j_l(k\,\chi) j_l (k , \chi^{\prime})
\langle {\Phi}(k,\chi ){\Phi}(k,\chi')\rangle \,.
\end{eqnarray}

In this expression there is a ``hidden'' projection effect, encoded in the geometric part containing $f(\chi)$, which will be responsible for an alteration in the position of the peaks; there is then an amplitude term, $(\phi^2/\phi^2_0 - 1)$, which can be safely regarded as slowly dependent of the redshift and is of order unity in all the models under study, allowing to gain an estimate of the size of effect as
\begin{equation}
\delta P_\kappa / P_\kappa \simeq - 128\,G\,\xi\,\phi^2_0 \;
\end{equation} 
thus, since for normalization reason the product $G \,\phi^2_0$ is always very close to one, the correction in the convergence power spectrum is sizeable even with values of the coupling parameter 
as small as $10^{-3}$, 
which is the typical value allowed by current experiments, $\omega_{JBD}=1/32\pi G\phi_{0}^{2}\xi^2 >4\cdot 10^{4}$ 
(\cite{bertotti_etal_2003}). 

\section{Conclusions}

We have shown how the weak gravitational lensing can be used as an investigation tool for the dynamics of dark energy or modified gravity. 
We have provided a systematic analytic treatment for the lensing observables in conjugation with the cosmological perturbations formalism, 
and we have checked the expected order of magnitude of the resulting effect in a popular model. \\
This formal part of the work is now being used in order to get a complete numerical sample in a wide variety of models, and we expect to be 
able to constrain different dark energy or modified gravity scenarios, by means of the accurate indicator represented by the magnetic modes of 
CMB polarization, within the next generation of lensing-devoted experiments. 

\section{Acknowledgements}
We warmly thank Martin White for valuable suggestions. V.A. is grateful to Andrew Liddle for support and many stimulating discussions during her stay at Sussex University, where part of this work has been written.

\end{document}